\documentclass[12pt]{article}
\usepackage{graphicx}
\def\beq{\begin{equation}}
\def\eeq{\end{equation}}
\def\bea{\begin{eqnarray}}
\def\eea{\end{eqnarray}}
\def\nn{\nonumber}
\def\sss{\scriptscriptstyle}
\def\lft{{\sss L}}
\def\rht{{\sss R}}
\def\roughly#1{\mathrel{\raise.3ex\hbox
{$#1$\kern-.75em\lower1ex\hbox{$\sim$}}}}

\def\epjc#1#2#3{{\it Eur.\ Phys.\ J.}\ {\bf C#1} (#2) #3}

\def\npb#1#2#3{{\it Nucl.\ Phys.} {\bf B#1} (#2) #3}
\def\plb#1#2#3{{\it Phys.\ Lett.} {\bf #1B} (#2) #3}

\def\prd#1#2#3{{\it Phys.\ Rev.} {\bf D#1} (#2) #3}
\def\newprd#1#2#3{{\it Phys.\ Rev.} {\bf D#1}: #3 (#2)}

\def\prl#1#2#3{{\it Phys.\ Rev.\ Lett.} {\bf #1} (#2) #3}

\begin{document}

\setlength{\baselineskip}{20pt}

\begin{flushright}
UdeM-GPP-TH-01-91 \\
\end{flushright}

\begin{center}
\bigskip

{\Large \bf Flavour-Changing Neutral Currents and Leptophobic $Z'$
Gauge Bosons} \\
\bigskip

Karine Leroux \footnote{leroux@lps.umontreal.ca} and
David London \footnote{london@lps.umontreal.ca}
\end{center}


\begin{center}
{\it Laboratoire Ren\'e J.-A. L\'evesque, Universit\'e de
Montr\'eal,}\\
{\it C.P. 6128, succ. centre-ville, Montr\'eal, QC, Canada H3C 3J7}
\end{center}

\begin{center}
 
\bigskip (\today)

\bigskip 

{\bf Abstract}

\end{center}

\begin{quote} 
Leptophobic $Z'$ gauge bosons can appear in models with an $E_6$ gauge
symmetry. We show that flavour-changing neutral currents can be
generated in some of these models due to the mixing of the ordinary
$d_\rht$, $s_\rht$ and $b_\rht$ quark fields with the exotic
$h_\rht$. Because the $Z'$ does not couple to charged leptons, the
constraints on the flavour-changing couplings $U^{\sss Z'}_{db}$ and
$U^{\sss Z'}_{sb}$ are relatively weak. Indeed, $B^0_q$--${\bar
B}^0_q$ mixing ($q=d,s$) can be dominated by $Z'$ exchange, which will
affect CP-violating rate asymmetries in $B$ decays. Rare hadronic $B$
decays can also be affected, while decays involving charged leptons
will be unchanged.
\end{quote}
\newpage

Many models of physics beyond the Standard Model (SM) predict the
existence of exotic fermions with non-canonical $SU(2)_\lft \times
U(1)_{\sss Y}$ quantum numbers, i.e.\ left-handed $SU(2)_\lft$
singlets and/or right-handed $SU(2)_\lft$ doublets. The ordinary SM
fermions can mix with these exotic fermions. It is well known that
such mixing may induce flavour-changing neutral currents (FCNC's)
\cite{LL}: if two ordinary quarks mix with the same exotic quark,
FCNC's are generated between the ordinary quarks. This FCNC is second
order in ordinary-exotic quark mixing.

This fact has been used to construct models with $Z$-mediated FCNC's
\cite{NirSilv}. Here one introduces an additional vector-singlet
charge $-1/3$ quark $h$, as is found in $E_6$ models, and allows it to
mix with the ordinary down-type quarks $d$, $s$ and $b$. Since the
weak isospin of the exotic quark is different from that of the
ordinary quarks, $Z$-mediated FCNC's among the ordinary down-type
quarks are induced. Note that it is only the mixing between the
left-handed components of the ordinary and exotic quarks which is
responsible for the FCNC: since $d_\rht$, $s_\rht$, $b_\rht$ and
$h_\rht$ all have the same $SU(2)_\lft \times U(1)_{\sss Y}$ quantum
numbers, their mixing cannot generate flavour-changing couplings of
the $Z$.

The $Z$-mediated FCNC couplings $U_{ds}^{\sss Z}$, $U_{db}^{\sss Z}$
and $U_{sb}^{\sss Z}$, which are in general complex, are constrained
by a variety of processes.  $U_{ds}^{\sss Z}$ is bounded by the
measurements of $\Delta M_{\sss K}$ ($K^0$--${\bar K}^0$ mixing),
$|\epsilon|$ (CP violation in the kaon system) and $K_\lft \to \mu^+
\mu^-$ \cite{NirSilv}, while the constraints on $U_{db}^{\sss Z}$ and
$U_{sb}^{\sss Z}$ come principally from the experimental limit on
$B(B\to \ell^+ \ell^- X)$ \cite{BELLE,bsll}. To the extent that the
constraints on $U_{db}^{\sss Z}$ and $U_{sb}^{\sss Z}$ allow
significant contributions to $B_q^0$--${\bar B}_q^0$ mixing ($q=d,s$),
CP asymmetries in $B$ decays may be affected by $Z$-mediated FCNC's
\cite{NirSilv,Bnewphysics}.

In general, models of new physics which contain exotic fermions also
predict the existence of additional neutral $Z'$ gauge bosons. The
same ideas which lead to $Z$-mediated FCNC's can be applied to the
$Z'$. That is, mixing among particles which have different $Z'$
quantum numbers will induce FCNC's due to $Z'$ exchange \cite{Z'FCNC}.
Surprisingly, these effects can be just as large as $Z$-mediated
FCNC's. Since the $U_{pq}^{\sss Z}$ are generated by mixings which
break weak isospin, they are expected to be at most $O(m/M)$, where
$m$ ($M$) is a typical light (heavy) fermion mass. On the other hand,
the $Z'$-mediated couplings $U_{pq}^{Z'}$ can be generated via mixings
of particles with the same weak isospin, and so they suffer no such
mass suppression.  Therefore, even though processes with $Z'$ exchange
are suppressed relative to those with $Z$ exchange by $M_{\sss
Z}^2/M_{\sss Z'}^2$, this is compensated by the fact that
$U_{pq}^{Z'}/U_{pq}^Z \sim M/m$.  Thus, the effects of $Z'$-mediated
FCNC's can be comparable to those of $Z$-mediated FCNC's.

In this paper we apply these ideas to leptophobic $Z'$ gauge bosons,
whose couplings to charged leptons vanish. Leptophobic $Z'$ bosons
were introduced several years ago in the context of the $R_b$--$R_c$
puzzle \cite{RbRc}, and as a possible explanation of anomalous
high-$E_{\sss T}$ jet events at CDF \cite{CDFjets}. Although these
experimental effects ultimately disappeared, thereby removing the
original motivation for such new physics, models with a leptophobic
$Z'$ still remain as viable candidates of physics beyond the SM, and
it is therefore worthwhile exploring their phenomenology.

In Ref.~\cite{Babu} it was shown that a leptophobic $Z'$ can appear in
$E_6$ models due to the mixing of the gauge boson kinetic terms. In
such models, if the $d_\rht$, $s_\rht$ and $b_\rht$ have different
$U(1)'$ quantum numbers than the $h_\rht$, then their mixing will
induce $Z'$-mediated FCNC's among the ordinary down-type quarks.
However, since the $Z'$ is leptophobic, these FCNC couplings will not
be constrained by limits on processes involving charged leptons, such
as $K_\lft \to \mu^+ \mu^-$ and $B\to \mu^+ \mu^- X$. Thus, the
constraints on such leptophobic $Z'$-mediated FCNC's may be
considerably weaker than those for $Z$-mediated FCNC's and, as a
consequence, there may be large effects in $B$ decays. These are the
issues which we examine in this paper.

We begin with a brief review of models with a leptophobic $Z'$ gauge
boson \cite{Rizzo}. We assume that the low-energy gauge symmetry is
$SU(2)_\lft \times U(1)_{\sss Y} \times U(1)'$, in which the $U(1)'$
arises from the breaking chain
\beq
E_6 \to SO(10) \times U(1)_\psi \to SU(5) \times U(1)_\chi \times
U(1)_\psi \to SU(2)_\lft \times U(1)_{\sss Y} \times U(1)' ~.
\eeq
Here, $U(1)'$ is a linear combination of $U(1)_\psi$ and $U(1)_\chi$,
with $Q' = Q_\psi \cos\theta - Q_\chi \sin\theta$, where $\theta$ is
the usual $E_6$ mixing angle. The fundamental representation of $E_6$
is a {\bf 27}, which decomposes under $SO(10)$ as a ${\bf 16} + {\bf
  10} + {\bf 1}$. The conventional embedding is to put all the
ordinary SM particles, along with a right-handed neutrino, into the
{\bf 16}. Within this embedding, the quantum numbers of all particles
are shown in Table~\ref{Qnostable}.

\begin{table*}[htbp]
\leavevmode
\begin{center}
\begin{tabular}{lcccc}
\hline
\hline
Particle & $SU(3)_c$ & $2\sqrt{6} Q_\psi$ & $2\sqrt{10} Q_\chi$ & $Y$ \\
\hline
$Q=(u,d)^T$   & {\bf 3}      &   1   & $-$1   & 1/6    \\
$L=(\nu,e)^T$ & {\bf 1}      &   1   &  3   &$-$1/2   \\
$u^c$ &$\overline{\mbox{\bf 3}}$       &   1   &  $-$1  & $-$2/3   \\
$d^c$ &$\overline{\mbox{\bf 3}}$       &   1   &  3   & 1/3    \\
$e^c$         &{\bf 1}       &   1   &  $-$1  &1     \\
$\nu^c$       &{\bf 1}       &   1   &  $-$5  &0     \\
$H=(N,E)^T$   &{\bf 1}       &  $-$2   &  $-$2  &$-$1/2  \\
$H^c=(N^c,E^c)^T$ &{\bf 1}   &  $-$2   &  2   &1/2   \\
$h$           & {\bf 3}      &  $-$2   &  2   &$-$1/3  \\
$h^c$ &$\overline{\mbox{\bf 3}}$       &  $-$2   &  $-$2  &   1/3  \\
$S^c$     &{\bf 1}           &   4   &  0   &0     \\
\hline
\hline
\end{tabular}
\caption{Quantum numbers of the particles contained in the {\bf 27} 
  representation of $E_6$ within the standard embedding. All fields
  are taken to be left-handed.}
\label{Qnostable}
\end{center}
\end{table*}

It is straightforward to show that if the $Z'$ coupling to fermions is
proportional to $Q'$, there is no value of $\theta$ which leads to
leptophobia (i.e.\ $Q'(L) = Q'(e^c) = 0$). However, the most general
$SU(2)_\lft \times U(1)_{\sss Y} \times U(1)'$-invariant lagrangian
includes a kinetic mixing term between the $U(1)_{\sss Y}$ and $U(1)'$
gauge bosons:
\beq
{\cal L}_{kin} = -{1\over {4}} W^a_{\mu\nu}W^{a\mu\nu} - {1\over {4}}
\tilde B^{\mu\nu}\tilde B_{\mu\nu} - {1\over {4}} \tilde
Z'^{\mu\nu}\tilde Z'_{\mu\nu} - {\sin \chi \over {2}}\tilde B_{\mu\nu}
\tilde Z'^{\mu\nu} ~,
\eeq
where the $W^a$, $\tilde B$ and $\tilde Z'$ represent the
$SU(2)_\lft$, $U(1)_{\sss Y}$ and $U(1)'$ fields. Due to the presence
of kinetic mixing, the physical $Z'$ can exhibit leptophobia. This can
be seen as follows \cite{Rizzo}.

The off-diagonal coupling of the ${\tilde B}$ and ${\tilde Z}'$ can be
removed by making the non-unitarity transformation
\bea
{\tilde B}_\mu & = & B_\mu - \tan\chi Z'_\mu ~, \nn\\
{\tilde Z}'_\mu & = & {Z'_\mu \over \cos\chi} ~.
\eea
With this transformation, the couplings of the physical gauge bosons
to fermions can be written as
\beq
{\cal L}_{int} = -\bar \psi \gamma^\mu [g T^a W^a_\mu + g'
Y_{\sss SM} B_\mu + g_{\sss Q'} (Q'+\sqrt {3\over {5}} \delta Y_{\sss
  SM}) Z'_\mu ] \psi ~,
\eeq
where $Q_{em}=T_{3L}+Y_{\sss SM}$ and $\delta\equiv -{\tilde g}_{\sss
  Y} \sin\chi / {\tilde g}_{\sss Q'}$ (${\tilde g}_{\sss Y}$ and
${\tilde g}_{\sss Q'}$ are, respectively, the $U(1)_{\sss Y}$ and
$U(1)'$ coupling constants). Assuming the couplings to be ``GUT''
normalized, the $Z'$-fermion interaction term can be written
\beq
{\cal L}(Z')_{int} = -\lambda {g \over \cos\theta_{\sss W}} \sqrt
{5 x_{\sss W}\over {3}} \, \bar \psi \gamma^\mu ( Q' + \sqrt{3\over
{5}} \delta Y_{\sss SM} ) \psi Z'_\mu ~,
\label{Z'ff}
\eeq
where $x_{\sss W} \equiv \sin^2 \theta_{\sss W}=e^2/g^2$ and
$\lambda=g_{\sss Q'}/g_{\sss Y}$. The key point here is that, due to
kinetic mixing, the $Z'$ coupling to fermions is no longer
proportional to $Q'$. It is this feature which leads to the
possibility of leptophobia.

The $Z'$-fermion coupling involves two unknown parameters: $\theta$
and $\delta$. Since leptophobia requires two couplings to vanish (the
$Z'$ coupling to $e^-$ and $e^+$), obviously this can be satisfied for
some choice of the two parameters. For example, for the conventional
embedding of Table~\ref{Qnostable}, one obtains a leptophobic $Z'$ for
$\tan\theta = \sqrt{3/5}$ and $\delta = -1/3$.

However, other embeddings are possible. First, since $L$ and $H$ have
the same $SU(2)_\lft \times U(1)_{\sss Y}$ quantum numbers, it is
always possible to switch the quantum numbers of $L$ and $H$ in
Table~\ref{Qnostable}. (This also holds for $d^c$ and $h^c$.) Second,
there are three states which are singlets under both $SU(3)_c$ and
$SU(2)_\lft$ -- the $e^c$ field can be assigned to any one of these
three. Of course, since not all three states have the same electric
charge, these three embeddings correspond to different definitions of
the charge generator. Thus, in addition to changing the $e^c$
assignment, one also has to change the assignment of the $u^c$
field. (Note that there are three states which are $SU(2)_\lft$
singlets and ${\bar{\bf 3}}$'s under $SU(3)_c$. The three $e^c$
embeddings are equivalent to assigning the $u^c$ to each of these
states.) There are thus a total of six choices for possible embeddings
of the charged leptons in the {\bf 27} representation of $E_6$. A
leptophobic $Z'$ can be produced for any of these.

We summarize the six embeddings, along with the corresponding values
of $\tan\theta$ and $\delta$ which produce leptophobia, in
Table~\ref{embeddings}. Note that the first four models have been
discussed in Ref.~\cite{Rizzo}, while the last two are new
possibilities.

\begin{table*}[htbp]
\leavevmode
\begin{center}
\begin{tabular}{ccccccccc}
\hline
\hline
Model & & $2\sqrt{6} Q_\psi$ & $2\sqrt{10} Q_\chi$ & &
$2\sqrt{6} Q_\psi$ & $2\sqrt{10} Q_\chi$ & $\tan\theta$ & $\delta$ \\
\hline
1 & $L$: & 1 & 3 & $e^c$: & 1 & $-1$ & $\sqrt{3/5}$ & $-1/3$ \\
2 & $L$: & $-2$ & $-2$ & $e^c$: & 1 & $-1$ & $\sqrt{3/5}$ & $-1/3$ \\
3 & $L$: & 1 & 3 & $e^c$: & 1 & $-5$ & $\sqrt{15}$ & $-\sqrt{10}/3$ \\
4 & $L$: & $-2$ & $-2$ & $e^c$: & 1 & $-5$ & $\sqrt{5/27}$ & $-\sqrt{5/12}$ \\
5 & $L$: & 1 & 3 & $e^c$: & 4 & 0 & $\sqrt{5/3}$ & $-\sqrt{5/12}$ \\
6 & $L$: & $-2$ & $-2$ & $e^c$: & 4 & 0 & 0 & $-\sqrt{10}/3$ \\
\hline
\hline
\end{tabular}
\caption{$Q_\psi$ and $Q_\chi$ quantum numbers of $L$ and $e^c$ for the 
  six embeddings of charged leptons in the {\bf 27} representation of
  $E_6$, along with the values of $\theta$ and $\delta$ which produce
  a leptophobic $Z'$ gauge boson.}
\label{embeddings}
\end{center}
\end{table*}

Now, what interests us is the possibility of $Z'$-mediated FCNC's.
This can occur if the $d$, $s$ and $b$ quarks mix with the $h$ quark.
However, as discussed earlier, mixing of the $d_\lft$, $s_\lft$ and
$b_\lft$ fields with $h_\lft$ will lead to $Z$-mediated FCNC's among
the ordinary fermions. These $Z$-mediated FCNC's will always dominate
over any $Z'$-mediated FCNC's induced by the mixing of the left-handed
fermions. On the other hand, if only the right-handed fields mix, then
no $Z$-mediated FCNC's will be generated, while $Z'$-mediated FCNC's
will appear if the $d_\rht$, $s_\rht$ and $b_\rht$ fields have
different $Z'$ quantum numbers than the $h_\rht$. The question then is
the following: of the six leptophobic models, are there any in which
$Q'(d^c) \ne Q'(h^c)$?

In order to answer this question, for each of the six models of
Table~\ref{embeddings}, we calculate the $U(1)'$ charges of the $d^c$
and $h^c$ fields using $Q' = Q_\psi \cos\theta - Q_\chi \sin\theta$.
(Note that, as discussed above, the labels `$d^c$' and `$h^c$' are
arbitrary: one can always switch the two fields.) The results are
shown in Table~\ref{dhcharges}. Of the six models, two of them ---
models 4 and 5 --- have $Q'(d^c) \ne Q'(h^c)$. Thus, in these models,
the mixing of $d_\rht$, $s_\rht$ and $b_\rht$ with the $h_\rht$ will
lead to FCNC's mediated by the exchange of a leptophobic $Z'$ gauge
boson.

\begin{table*}[htbp]
\leavevmode
\begin{center}
\begin{tabular}{ccccccccc}
\hline
\hline
Model & & $2\sqrt{6} Q_\psi$ & $2\sqrt{10} Q_\chi$ & $Q'$ & &
$2\sqrt{6} Q_\psi$ & $2\sqrt{10} Q_\chi$ & $Q'$ \\
\hline
1 & $d^c$: & 1 & 3 & $-1/2\sqrt{15}$ & $h^c$: & $-2$ & $-2$ & $-1/2\sqrt{15}$ \\
2 & $d^c$: & 1 & 3 & $-1/2\sqrt{15}$ & $h^c$: & $-2$ & $-2$ & $-1/2\sqrt{15}$ \\
3 & $d^c$: & 1 & $-1$ & $1/2\sqrt{6}$ & $h^c$: & $-2$ & $-2$ & $1/2\sqrt{6}$ \\
4 & $d^c$: & 1 & $-1$ & $1/4$ & $h^c$: & $-2$ & $-2$ & $-1/4$ \\
5 & $d^c$: & 1 & $-1$ & $1/4$ & $h^c$: & 1 & 3 & $-1/4$ \\
6 & $d^c$: & 1 & $-1$ & $1/2\sqrt{6}$ & $h^c$: & 1 & 3 & $1/2\sqrt{6}$ \\
\hline
\hline
\end{tabular}
\caption{$U(1)'$ quantum numbers of $d^c$ and $h^c$ for each of the
six models given in Table~\protect\ref{embeddings}, calculated using
$Q' = Q_\psi \cos\theta - Q_\chi \sin\theta$.}
\label{dhcharges}
\end{center}
\end{table*}

In order to examine the constraints on such $Z'$-mediated FCNC
couplings, we parametrize them as
\beq 
{\cal L}^{\sss Z'}_{\sss FCNC} = - {g \over 2 \cos\theta_{\sss W}} \,
U_{qp}^{\sss Z'} \, {\bar d}_{q\rht} \gamma^\mu d_{p\rht} Z'_\mu ~.
\eeq
As is the case for $Z$-mediated FCNC's, the coupling $U^{\sss
Z'}_{ds}$ is strongly constrained by measurements of $\Delta M_{\sss
K}$ and $|\epsilon|$ in the kaon system \cite{NirSilv}:
\bea
\left| {\rm Re}\left( U_{ds}^{\sss Z'} \right)^2 \right| {M_{\sss
Z}^2 \over M_{\sss Z'}^2} & \le & 4.1 \times 10^{-7} ~~~~~~~~~
(\Delta M_{\sss K}) ~, \nn\\
\left| {\rm Im}\left( U_{ds}^{\sss Z'} \right)^2 \right| {M_{\sss
Z}^2 \over M_{\sss Z'}^2} & \le & 2.6 \times 10^{-9} ~~~~~~~~~
(|\epsilon|) ~.
\eea

Note that, unlike the flavour-changing coupings of the $Z$, there are
no constraints on $U_{ds}^{\sss Z'}$ from $K_\lft \to \mu^+ \mu^-$
since the leptophobic $Z'$ does not couple to charged leptons.
Similarly, the couplings $U^{\sss Z'}_{db}$ and $U^{\sss Z'}_{sb}$ are
unconstrained by the experimental limit on $B(B\to \mu^+ \mu^- X)$,
which is the main constraint on the flavour-changing couplings of the
$Z$ to the $b$ quark. On the other hand, $Z'$-mediated FCNC's do
contribute to the process $b \to s \nu{\bar \nu}$ \cite{bsnunu}, for
which ALEPH has an experimental limit \cite{ALEPH}:
\beq
B(b \to s \nu {\bar \nu}) \le 6.4 \times 10^{-4} ~.
\eeq
In order to compute the contribution of the $U^{\sss Z'}_{sb}$
coupling to this process, we need the coupling of the $Z'$ to $\nu
{\bar \nu}$. Since the $Z'$ is leptophobic, it does not couple to
$L_\lft$, which includes both $e^-_\lft$ and $\nu_{e\lft}$. However,
it does couple to the right-handed neutrino, and this must be taken
into account.

In $E_6$, there are two candidates for the right-handed neutrino: the
fields labelled $\nu^c$ and $S^c$ in Table~\ref{Qnostable}. In
Table~\ref{nuccharges} we present the $U(1)'$ charges of these two
fields. (Note that, as before for $L$/$H$ and $d^c$/$h^c$, one can
always exchange the fields $\nu^c \leftrightarrow S^c$, so that the
labels are arbitrary.) From this table we see that the leptophobic
$Z'$ does indeed couple to the right-handed neutrino, and so it can
contribute to $b \to s \nu {\bar \nu}$\footnote{Of course, it could be
that the light right-handed neutrino is that linear combination of
$\nu^c$ and $S^c$ whose coupling to the $Z'$ vanishes, in which case
there are no $Z'$ contributions to $b \to s \nu {\bar \nu}$. Although
logically possible, we do not consider this fine-tuned solution
here.}.

\begin{table*}[htbp]
\leavevmode
\begin{center}
\begin{tabular}{ccccccccc}
\hline
\hline
Model & & $2\sqrt{6} Q_\psi$ & $2\sqrt{10} Q_\chi$ & $Q'$ & &
$2\sqrt{6} Q_\psi$ & $2\sqrt{10} Q_\chi$ & $Q'$ \\
\hline
4 & $\nu^c$: & 1 & $-1$ & $1/4$ & $S^c$: & 4 & 0 & $3/4$ \\
5 & $\nu^c$: & 1 & $-1$ & $1/4$ & $S^c$: & 1 & $-5$ & $3/4$ \\
\hline
\hline
\end{tabular}
\caption{$U(1)'$ quantum numbers of $\nu^c$ and $S^c$ for models 4 and
5 of Table~\protect\ref{embeddings}, calculated using $Q' = Q_\psi
\cos\theta - Q_\chi \sin\theta$.}
\label{nuccharges}
\end{center}
\end{table*}

Taking $Q'(\nu^c) = 1/4$, the coupling of the $Z'$ to the right-handed
neutrino can be written as
\beq
- {g \over 2 \cos\theta_{\sss W}} Q^{\sss Z'}_{\nu_\rht}
{\bar\nu}_\rht \gamma^\mu \nu_\rht Z'_\mu ~,
\eeq
with [see Eq.~(\ref{Z'ff})]
\beq 
Q^{\sss Z'}_{\nu_\rht} = {1\over 2} \lambda \sqrt{5 \sin^2\theta_{\sss
W} \over 3} = 0.31 ~,
\eeq
where we have taken $\lambda = 1$ (its precise value depends on the
details of unification). The contribution of $Z'$-mediated FCNC's to
$b \to s \nu {\bar \nu}$ is then given by
\beq
{B(B \to X_s \nu {\bar\nu}) \over B(B \to \mu \nu X)} = { \left(
Q^{\sss Z'}_{\nu_\rht} \right)^2 \left| U^{\sss Z'}_{sb} \right|^2
\over |V_{ub}|^2 + F_{ps} |V_{cb}|^2 } \left( {M_{\sss Z}^2 \over
M_{\sss Z'}^2} \right)^2 ~,
\eeq
where $F_{ps} \simeq 0.5$ is a phase-space factor. This yields the
constraint
\beq
\left| U^{\sss Z'}_{sb} \right| {M_{\sss Z}^2 \over M_{\sss Z'}^2} \le
7.1 \times 10^{-3} ~.
\label{Z'constraint}
\eeq
This can be turned into a bound on $U^{\sss Z'}_{sb}$ if one assumes a
value for $M_{\sss Z'}$. The only experimental constraint on
leptophobic $Z'$ gauge bosons comes from the D0 experiment \cite{D0},
which excludes the mass range $365~{\rm GeV} \le M_{\sss Z'} \le
615~{\rm GeV}$ for a $Z'$ with quark couplings equal to those of the
$Z$. (Interestingly, light leptophobic $Z'$ bosons are not ruled out.)

There are two points to be stressed here. First, the constraints on
$Z'$-mediated $b \to s$ transitions are quite a bit weaker than those
on the corresponding $Z$-mediated FCNC's. The most recent result from
BELLE gives \cite{BELLE}
\beq
B(B \to X_s e^+ e^-) \le 1.01 \times 10^{-5} ~,
\label{BELLEbound}
\eeq
which leads to the constraint
\beq
\left\vert U_{sb}^{\sss Z} \right\vert \le 7.6 \times 10^{-4} ~.
\label{ZFCNCconstraint}
\eeq
This is about an order of magnitude more stringent than the
corresponding $Z'$ FCNC constraint of Eq.~(\ref{Z'constraint}). Thus,
effects due to leptophobic $Z'$-mediated FCNC's in $b\to s$ processes
may be larger than those due to $Z$-mediated FCNC's.

Second, unlike $Z$-mediated FCNC's, there are no constraints on $b\to
d$ transitions from $B$ decays. Thus, here too the effects of
$Z'$-mediated FCNC's can be considerably larger than those due to $Z$
exchange.

Of course, $Z'$-mediated FCNC's will contribute to $B^0_q$--${\bar
B}^0_q$ mixing ($q=d,s$):
\beq
M_{12}^{\sss Z'} (B_q) = {\sqrt{2} G_{\sss F} M_{\sss B_q} \eta_{\sss
B_q} \over 12} \, {M_{\sss Z}^2 \over M_{\sss Z'}^2} \, f_{\sss B_q}^2
B_{\sss B_q} (U_{qb}^*)^2 ~.
\eeq
These contributions can be compared with those of the SM:
\beq
{\Delta M_q^{\sss Z'} \over \Delta M_q^{\sss W} } = {\sqrt{2} \pi^2
\over G_{\sss F} M_{\sss W}^2} \, { 1 \over x_t f_2(x_t) } \, {M_{\sss Z}^2
\over M_{\sss Z'}^2} \, { \left\vert U_{qb}^{\sss Z'} \right\vert^2
\over \left\vert V_{tq}V_{tb} \right\vert^2 } = 80 \, {M_{\sss Z}^2
\over M_{\sss Z'}^2} \, { \left\vert U_{qb}^{\sss Z'} \right\vert^2
\over \left\vert V_{tq} \right\vert^2 } ~,
\eeq
where we have taken $\left\vert V_{tb} \right\vert = 1$ and $m_t =
170$ GeV. 

Consider first $B_s^0$--${\bar B}_s^0$ mixing. As a figure of merit,
we assume that $U^{\sss Z'}_{sb} = 0.1$ and $M_{\sss Z'} = 750$ GeV,
which satisfy the bound of Eq.~(\ref{Z'constraint}). Taking $|V_{ts}|
= |V_{cb}| = 0.04$, this yields
\beq
{\Delta M_s^{\sss Z'} \over \Delta M_s^{\sss W} } = 7.2 ~.
\eeq
Thus, $B_s^0$--${\bar B}_s^0$ mixing can be completely dominated by
the exchange of a leptophobic $Z'$. This is in stark contrast to
$Z$-mediated FCNC's. With the constraint of
Eq.~(\ref{ZFCNCconstraint}), we have
\beq
{\Delta M_s^{\sss Z} \over \Delta M_s^{\sss W} } = 
80 \, { \left\vert U_{sb}^{\sss Z} \right\vert^2
\over \left\vert V_{ts} \right\vert^2 } \le 0.029 ~.
\eeq
The contribution of $Z$-mediated FCNC's to $B_s^0$--${\bar B}_s^0$
mixing is therefore negligible compared to that of the SM.

Turning to $B_d^0$--${\bar B}_d^0$ mixing, since there are no
constraints on $U_{db}^{\sss Z'}$ from $B$ decays, it is clear that
this mixing can also be dominated by $Z'$-mediated FCNC's. (Note that
it was shown in Ref.~\cite{Bnewphysics} that $B_d^0$--${\bar B}_d^0$
mixing could be dominated by $Z$-mediated FCNC's. Since the bounds on
$U_{db}^{\sss Z}$ have not changed, this still holds true. Even if it
is assumed that the bound of Eq.~(\ref{BELLEbound}) also applies to
$b\to d e^+e^-$, in which case $\left\vert U^{\sss Z}_{db} \right\vert
< 7.6 \times 10^{-4}$, the contributions of $Z$-mediated FCNC's to
$B_d^0$--${\bar B}_d^0$ mixing can still be important.)

{}From the above, we have seen that both $B^0_d$--${\bar B}^0_d$ and
$B^0_s$--${\bar B}^0_s$ mixing can be dominated by $Z'$-mediated
FCNC's. Since the couplings $U^{\sss Z'}_{qb}$ ($q=d,s$) can be
complex, there may be large effects on CP-violating rate asymmetries
in $B$ decays. Furthermore, if $B^0_q$--${\bar B}^0_q$ mixing is
significantly affected by this type of new physics, one expects that
rare $B$ decays will also be affected \cite{Bnewphysics}. What
distinguishes leptophobic $Z'$-mediated FCNC's from other models of
new physics is that its effects will only show up in rare hadronic $B$
decays; leptonic decays such as $b\to q \ell^+ \ell^-$ and $B_q^0 \to
\ell^+\ell^-$ will be unaffected. This provides a rather unique
``smoking-gun'' signal for this type of new physics.

To sum up: leptophobic $Z'$ gauge bosons can appear in models with an
$E_6$ gauge symmetry due to mixing of the gauge-boson kinetic
terms. There are a total of six fermion embeddings in the {\bf 27} of
$E_6$ which can produce leptophobia. Of these, we have shown that
flavour-changing neutral currents (FCNC's) can be generated in two of
these models. This is due to the mixing of the right-handed components
of the ordinary $d$, $s$ and $b$ quarks with the exotic $h$ quark.
Since all of these particles have the same weak isospin, this mixing
can be quite large.

The flavour-changing coupling $U^{\sss Z'}_{ds}$ is strongly
constrained by measurements of $\Delta M_{\sss K}$ and $|\epsilon|$ in
the kaon system. However, because the $Z'$ does not couple to charged
leptons, the constraints on $U^{\sss Z'}_{db}$ and $U^{\sss Z'}_{sb}$
are relatively weak -- they are bounded only by the experimental limit
on $B(b \to s \nu {\bar \nu})$. (This is in contrast to $Z$-mediated
FCNC's. For these, the constraints on the $U^{\sss Z}_{qb}$ ($q=d,s$)
from $B(B\to X\ell^+\ell^-)$ are quite stringent.) The result is that
both $B^0_d$--${\bar B}^0_d$ and $B^0_s$--${\bar B}^0_s$ mixing can be
dominated by $Z'$-mediated FCNC's. If there are significant
new-physics effects due to $Z'$ exchange in $B^0_q$--${\bar B}^0_q$
mixing, this will affect CP-violating rate asymmetries in $B$
decays. In addition, one expects that rare hadronic $B$ decays will
also be affected. The fact that only hadronic decays are affected, and
not leptonic decays such as $b\to q \ell^+ \ell^-$ and $B_q^0 \to
\ell^+\ell^-$, provides a unique signal for leptophobic $Z'$-mediated
FCNC's.

\section*{\bf Acknowledgments}

D.L. thanks E. Nardi and T. Rizzo for helpful communications. This
work was financially supported by NSERC of Canada and les Fonds FCAR
du Qu\'ebec.


\begin{thebibliography}{99}
  
\bibitem{LL} P. Langacker and D. London, \prd{38}{1988}{886}.

\bibitem{NirSilv} Y. Nir and D. Silverman, \prd{42}{1990}{1477}.

\bibitem{BELLE} K. Abe et al., (BELLE Collaboration), hep-ex/0107072.

\bibitem{bsll} S. Glenn et al., (CLEO Collaboration),
\prl{80}{1998}{2289}; B. Abbott et al., (D0 Collaboration),
\plb{423}{1998}{419}; C. Albajar et al., (UA1 Collaboration),
\plb{262}{91}{163};

\bibitem{Bnewphysics} M. Gronau and D. London, \prd{55}{1997}{2845}.
 
\bibitem{Z'FCNC} E. Nardi, \prd{48}{1993}{1240}; J. Bernab\'eu,
E. Nardi and D. Tommasini, \npb{409}{1993}{69}.
  
\bibitem{RbRc} The LEP Collaborations and the LEP Electroweak Working
Group, CERN-PPE/95/172.
  
\bibitem{CDFjets} F. Abe et al., CDF Collaboration, \prl{77}{1996}{438}.
  
\bibitem{Babu} K.S. Babu, C. Kolda and J. March-Russell,
\prd{54}{1996}{4635}, \prd{57}{1998}{6788}.
  
\bibitem{Rizzo} For a review of $Z'$ leptophobia, see T. Rizzo,
\newprd{59}{1999}{015020}.

\bibitem{bsnunu} Y. Grossman, Z. Ligeti and E. Nardi,
\npb{465}{1996}{753}, (Erratum) \npb{480}{1996}{753}.

\bibitem{ALEPH} R. Barate et al., ALEPH Collaboration,
\epjc{19}{2001}{213}.

\bibitem{D0} B. Abbott et al., D0 Collaboration,
Fermilab-Conf-97-356-E, presented at 18th International Symposium on
Lepton - Photon Interactions (LP 97), Hamburg, Germany, 28 Jul - 1 Aug
1997, and presented at International Europhysics Conference on
High-Energy Physics (HEP 97), Jerusalem, Israel, 19-26 Aug 1997.

\end{thebibliography}
\end{document}